\documentclass[aps,11pt]{revtex4}
\usepackage{graphicx}

\begin{document}

\title{A fast high-order method to calculate wakefield forces in
an electron beam}

\author{Ji Qiang, Chad Mitchell, Robert D. Ryne
}

\affiliation{
Lawrence Berkeley National Laboratory,
          1 Cyclotron Road, Berkeley, CA 94720 \\
}

\begin{abstract}
In this paper we report on a high-order fast method 
to numerically calculate
wakefield forces in an electron beam given a wake function model. 
This method
is based on a Newton-Cotes quadrature rule for integral approximation 
and an FFT method for discrete summation that results in 
an $O(Nlog(N))$ computational cost, where $N$ is the number of grid points. 
Using the Simpson quadrature rule with an accuracy of $O(h^4)$, 
where $h$ is the grid size, we present numerical calculation of the
wakefields from a resonator wake function model and from a one-dimensional
coherent synchrotron radiation (CSR) wake model. Besides the
fast speed and high numerical accuracy, the 
calculation using the direct line density instead of the first derivative of
the line density avoids numerical filtering of the electron density function for computing the CSR wakefield force.
\end{abstract}

\maketitle

\section{Introduction}

The wakefield forces from the interaction of electrons 
with external materials and from the interactions of electrons inside the
beam through radiation have significant effects on the electron beam quality
for next generation light sources.
Those wakefields can cause the loss of beam energy, 
increase the beam energy spread,
increase beam emittance through the magnetic bunch compressor,
and drive microbunching instability through the linac.
Including the wakefield effects in a beam dynamics tracking code
is crucial for designing a beam delivery system for next generation
light sources and has been included in a number of tracking 
codes~\cite{qiang0,elegant,lietrack,qiang00}.

The numerical calculation of the wakefield forces on electrons involves 
calculating a convolution between the wake function and the charge
density of the beam.
The direct numerical calculation of the convolution for the wakefield
has a computational cost scaling as $O(N^2)$, where $N$ is the number of
grid points in the computational domain. 
Fortunately, such a discretized convolution summation on
a uniform grid can be calculated using a cyclic summation on a doubled 
computational domain using an FFT based method~\cite{hockney,nr,qiang3,ryne1}. 
This reduces the computational
cost from $O(N^2)$ to $O(Nlog(N))$. 
In our previous study, we have developed a high-order FFT based method
to numerically calculate the convolution with a smooth kernel~\cite{qiang1}. 
In this paper, we extend the previous study by applying a high-order
Simpson quadrature rule to the wakefield convolution with finite 
discontinuity of the kernel. We then apply this method to the calculation of
the wakefields from an external resonator wake function (an approximation
to the AC resistance wall wake function) and from
a steady-state CSR wake function.

The organization of the paper is as follows: after a brief introduction, we
present the computational method in Section 2, application examples
to calculate the wakefield forces from an external resonator wake function 
and from
a steady-state CSR wake function in Section 3, and final summary in Section 4.

\section{Computational Method}

We consider the longitudinal wakefield on an electron at longitudinal
location $z$ inside 
an electron beam can
be written as
\begin{eqnarray}
E(z) & = & \int_{-\infty}^{\infty} W(z-z') \rho(z') dz'
\end{eqnarray}
where $W$ is the wake function from the interaction of electrons 
with external materials such as 
resistive wall wake or the interaction among electrons themselves 
such as the coherent synchrotron radiation (CSR) inside a bending
magnet, and $\rho$ is the beam line density distribution.
For the external wake function, the variable $z$ stands for 
$ct$, where $t$ is the time of flight difference with respect to
the head of the beam, and $c$ is the speed of light
in vacuum. For the CSR wake function, the variable $z$ stands 
for the longitudinal spatial position.
The above convolution integral can be written as the summation of $M$ equal subinterval integrals:
\begin{eqnarray}
E(z) & = & \sum_{i=1}^{M} \int_{z'_i}^{z'_{i+1}} W(z-z') \rho(z') dz'
\end{eqnarray}
where $z'_i = (i-1)(z/M), i = 1,\cdots,M$.
For the integral between $z'_i$ and $z'_{i+1}$, the closed Newton-Cotes formula of degree $k$ 
at $k+1$ equally spaced points can be written as
\begin{eqnarray}
\int_{z'_i}^{z'_{i+1}} W(z-z') \rho(z') dz' & \approx & \sum_{j=0}^{k} w_j W(z-z'_j) \rho(z'_j) 
\end{eqnarray}
where $z'_j=z'_i + j dz$, $dz = z/(kM)$, and the weight $w_j$ is associated with
integral of the $j_{th}$ Lagrange basis polynomial.
For $k=1$, this is known as the trapezoidal quadrature rule; for $k=2$, this is the Simpson rule;
for $k=3$, Simpson's $3/8$ rule; for $k=4$, Boole's rule, etc~\cite{math,math2}. 
Using the extended three-point Simpson rule on a uniform grid, the above convolution integral can be approximated as
\begin{eqnarray}
E(z) & = & \bar{E}(z) + O(h^4)
\end{eqnarray}
where $h$ is the grid size, and $\bar{E}(z)$ is the numerical approximation
of the convolution integral.  
For a grid point $z_i$, if $i$ is an odd number, the Simpson quadrature rule
yields
\begin{eqnarray}
\bar{E}(z_i) & = & \frac{1}{3} h (f_1 + 4 \sum_{j=1}^{(i-1)/2} f_{2j} + 2 \sum_{j=2}^{(i-1)/2} f_{2j-1}+f_i) 
\end{eqnarray}
if $i$ is an even number, it yields
\begin{eqnarray}
\bar{E}(z_i) & = & \frac{1}{3} h (3 f_1/2 + 5 f_2/2 + 2 \sum_{j=2}^{(i-2)/2} f_{2j} + 4 \sum_{j=2}^{i/2} f_{2j-1}+f_i) 
\end{eqnarray}
where the kernel function $f$ is given by
\begin{eqnarray}
f(z,z') & = & W(z-z') \rho(z')
\end{eqnarray} 
and $f_j(z_i) = f(z_i,z'_j)$ with $z'_j = (j-1)h$.
The above numerical summation 
can be rewritten as
\begin{eqnarray}
\bar{E}(z_i) & = & \frac{1}{3} h \sum_{j=1}^{N} \bar{W}(z_i-z_j) \bar{\rho}(z_j) + \frac{1}{3} h W(0) \rho(z_i)
\end{eqnarray}
with
\begin{eqnarray}
\bar{\rho}(z_j) & = & \left\{ \begin{array}{r@{\quad:\quad}l}
{\rho}(z_j) & j = 1 \\
4 {\rho}(z_j) & j = 2 l \\
2 \rho(z_j)    & j = 2 l - 1 
                              \end{array}
		      \right.  
\end{eqnarray}
for odd grid number $i$, and
\begin{eqnarray}
\bar{\rho}(z_j) & = & \left\{ \begin{array}{r@{\quad:\quad}l}
3 {\rho}(z_j)/2 & j = 1 \\
5 {\rho}(z_j)/2 & j = 2 \\
2 {\rho}(z_j) & j = 2 l \\
4 \rho(z_j)    & j = 2 l - 1 
                              \end{array}
		      \right.  
\end{eqnarray}
for even grid number $i$, and
the modified wake function $\bar{W}$ is given by
\begin{eqnarray}
\bar{W}(z) & = & \left\{ \begin{array}{r@{\quad:\quad}l}
{W}(z) & z > 0 \\
0 & z \leq 0 
                              \end{array}
		      \right.  
\end{eqnarray}

The direct calculation of the above summation requires $O(N)$ operations for a
single point $z_i$, where $i=1,\cdots,N$. 
To obtain the convolution for all $N$ points on the grid,
the total computational cost will be $O(N^2)$. 
Fortunately, the above convolution summation 
can 
be replaced by a cyclic summation in the
double-gridded computational domain:
\begin{eqnarray}
\bar{E}_{c}(z_i) & = & \frac{1}{3} h \sum_{j=1}^{2N} \bar{W}_c(z_i-z_j) \bar{\rho}_c(z_j) 
\end{eqnarray}
where $i=1,\cdots,2N$, \ $j=1,\cdots,2N$ and
\begin{eqnarray}
\bar{\rho}_c(z_j) & = & \left\{ \begin{array}{r@{\quad:\quad}l}
\bar{\rho}(z_j) & 1\le j \le N \\
 0            & N < j \le 2 N 
                              \end{array}
		      \right.  \\
\bar{W}_c(z_k) & = & \left\{ \begin{array}{r@{\quad:\quad}l}
\bar{W}(z_k) & 1\le k \le N;  \\
\bar{W}(-z_{2N+2-k})   & N< k \le 2 N;  \\
                              \end{array}
		      \right.  
\end{eqnarray}
where the new density function $\bar{\rho}_c(z)$ and the new wake function 
$\bar{W}_c(z)$ have the properties:
\begin{eqnarray}
\bar{\rho}_c(z) & = & \bar{\rho}_c(z+2(L+h)) \\
\bar{W}_c(z) & = & \bar{W}_c(z+2(L+h)) 
\end{eqnarray}
where $L = (N-1)h$. From above definition, one can show that
the cyclic summation gives the same value as the convolution summation 
within the original domain, i.e.
\begin{eqnarray}
\bar{E}(z_i) & = & \bar{E}_c(z_i) \hspace{1cm} for~ \ i=1, \ N.
\end{eqnarray}
In the cyclic summation, the kernel is a discrete periodic function
in the doubled computational domain.
This cyclic summation can be calculated using the FFT method, i.e.
\begin{eqnarray}
\bar{E}_{c}(z_i) & = & \it{InvFFT}(\bar{E}_{c}(\omega))  
\end{eqnarray}
Here, $\it{InvFFT}$ denotes the inverse FFT of the function $\bar{E}_{c}(\omega)$
that is given by
\begin{eqnarray}
\bar{E}_{c}(\omega) & = & \bar{W}_c(\omega) \ \bar{\rho}_c(\omega)
\end{eqnarray}
where $\bar{W}_c(\omega)$ and $\bar{\rho}_c(\omega)$ denote the forward FFT of the function
$\bar{W}_c$ and the $\bar{\rho}_c$ respectively. 
The computational operations required to 
calculate cyclic summation using the above FFT method
is of $O(Nlog(N)$. 
\section{Application Examples}

\begin{figure}[htb]
\centering
\includegraphics*[angle=270,width=75mm]{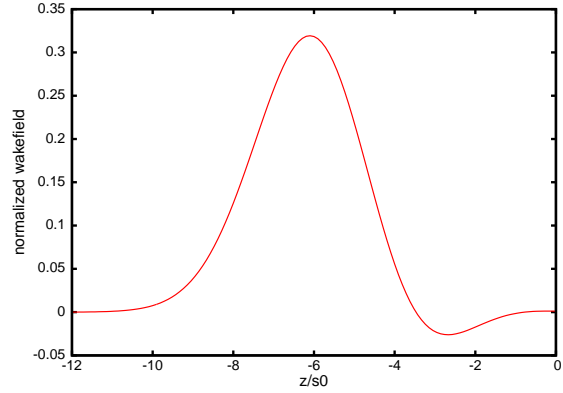}
\caption{The wakefield force as a function of 
bunch length coordinate in an electron beam inside
a Copper pipe with resonator wake function.
}
\label{fig1}
\end{figure}

\begin{figure}[htb]
\centering
\includegraphics*[angle=270,width=75mm]{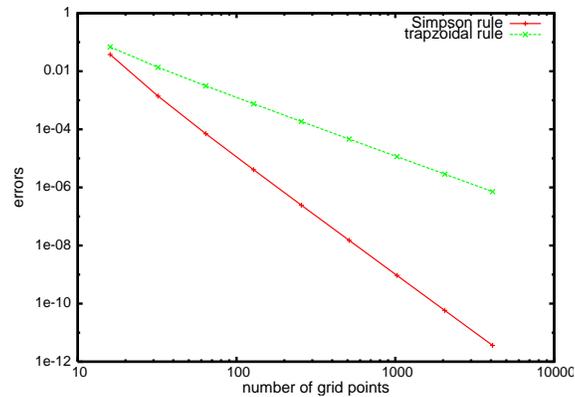}
\caption{The errors at the center of the electron beam 
as a function grid points. The red line is from the Simpson
quadrature rule and the green line is from the trapzoidal rule.}
\label{figrwkerr}
\end{figure}
As an illustration, 
we use the above method to calculate the wakefields inside an electron 
beam from a resonator wake function and from the CSR wake
function.
The resonator wake function is given as~\cite{bane}
\begin{eqnarray}
W(s) & = & \frac{4}{a^2} \exp(-(s/s_0)/\Gamma) \cos((8/\Gamma)^{1/4}(s/s_0)); \ for \ \ s \ge 0
\end{eqnarray}
where $s=z-z'$ is the separation between the observation point and
the source point, $s_0=(2a/(Z_0 \sigma))^{1/3}$ is the characteristic 
length of the wake, $a$ is the radius of the pipe, $\sigma$ is the conductivity
of the pipe material, $Z_0$ is the vacuum impedance,
$\Gamma = c \tau/s_0$ is the normalized relaxation time, 
and $\tau$ is the relaxation time.
This wake function provides a good approximation to the AC
resistive wall wake function at large $\Gamma$~\cite{bane}.

Figure~\ref{fig1} shows the normalized wakefield in an electron beam
moving inside a Copper pipe ($\Gamma=1$) computed using the above algorithm.
The beam has an rms bunch length of $1.5 s_0$ 
and a Gaussian current distribution.
The electron
beam loses energy to the wall due to the resistance of the conducting pipe.
To verify the accuracy of the above algorithm, 
Figure~\ref{figrwkerr} shows the errors of the calculated 
wakefield at the center of the bunch as a function of
the number of grid points. 
As a comparison, we also
give the errors using the trapezoidal quadrature rule. 
As expected, the
above Simpson rule method converges much faster than the
trapezoidal rule based method. The accuracy of the 
Simpson rule calculated wakefield increases by more than
an order of magnitude as the number of grid points doubles. 

\begin{figure}[htb]
\centering
\includegraphics*[angle=270,width=75mm]{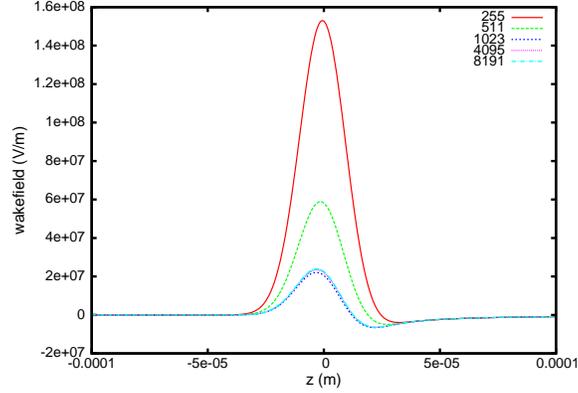}
\caption{
The CSR wakefield as a function of 
bunch length coordinate in an electron beam.}
\label{figcsrwkfld}
\end{figure}
Another application example of above numerical method is to 
calculate the wakefield force
from the one-dimensional steady state CSR wake.
The one-dimensional CSR wake function is given as~\cite{saldin}
\begin{eqnarray}
W(u(s)) & = & \frac{1}{4 \pi \epsilon}\frac{\gamma^4}{R^2}\{\frac{u^2/4-1}{2(1+u^2/4)^3}+\frac{1/6-u^2/18
- u^4/64}{(1+u^2/4)^3(1+u^2/12)^2)}\}  
\end{eqnarray}
and
\begin{eqnarray}
u(s) & = & (\sqrt{64+144\bar{s}^2}+12\bar{s})^{1/3}- (\sqrt{64+144\bar{s}^2}-12 \bar{s})^{1/3} 
\end{eqnarray}
where $\bar{s}=\frac{s\gamma^3}{R}$ is the normalized longitudinal separation between
the source point and the observation point, $R$ is the radius of the bending
magnet, and $\gamma$ is the relativistic factor of the electron beam.
The direct application of the above method with different number of grid points
is shown in Fig.~\ref{figcsrwkfld} for a $1$ nC Gaussian beam with $10$ micron rms bunch length at
$100$ MeV. It is seen that a few thousand grid points
are needed in order to obtain a numerically converged wakefield.
\begin{figure}[htb]
\centering
\includegraphics*[angle=270,width=75mm]{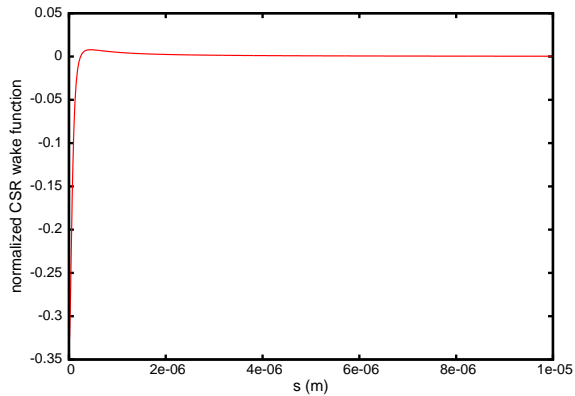}
\caption{The steady-state CSR wake function as a function of separation between
the observation point and the source point.}
\label{figcsrwake}
\end{figure}

\begin{figure}[htb]
\centering
\includegraphics*[angle=270,width=75mm]{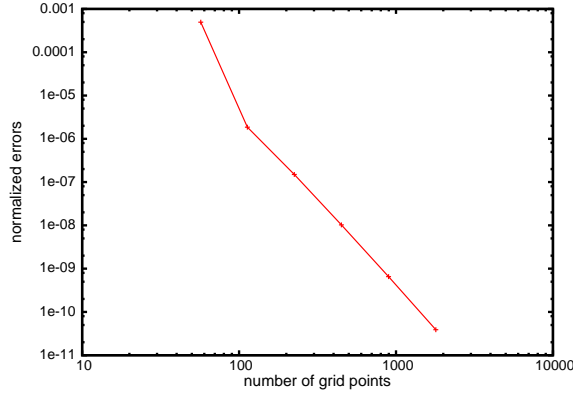}
\caption{The errors at the center of the bunch as a function of grid points for the first integral in Eq.~\ref{csr} using the Simpson rule. }
\label{figerr2}
\end{figure}

This large number of grid points is needed due to the sharp variation of the 
CSR wake function around
$s=0$ as shown in Fig.~\ref{figcsrwake}. 
It is seen that outside this short range (about
a few micron in this example) around zero,
the CSR wake function is much smoother and can be approximated by an
asymptotic function $s^{-4/3}$.
Recently, an integrated Green function method with linear basis was proposed
to treat the CSR wake function including the short range effect~\cite{ryne}. 
Here, we propose a modified Simpson rule method by
separating the range of
the original CSR wake function into two regions: 
one small region around the zero covering the short range CSR interaction, 
and the remainning region covering the long range interaction. 
The integral for the CSR wakefield can be written as:
\begin{eqnarray}
E(z) & = & \int_{-\infty}^{z-\delta z} W(z-z') \rho(z') dz' +
\int_{z-\delta z}^z W(z-z') \rho(z') dz'
\label{csr}
\end{eqnarray}
where $\delta z$ is a small distance around the zero to separate
the long-range CSR interaction from the short-range interaction.
The first integral in the above equation can be calculated using a 
Simpson quadrature
rule as shown in the previous section. 
Using the parameters in the above example with $\delta z = 3.2 \ um$ and
the Simpson quadrature rule, we calculated the errors 
at the center of the bunch as a function of the number of grid points 
for the first integral in Eq.~\ref{csr}.
The results are shown in Fig.~\ref{figerr2}.
As expected, the error of the integral decreases by more than an order
of magnitude as the number of the grid points doubles.

\begin{figure}[htb]
\centering
\includegraphics*[angle=270,width=75mm]{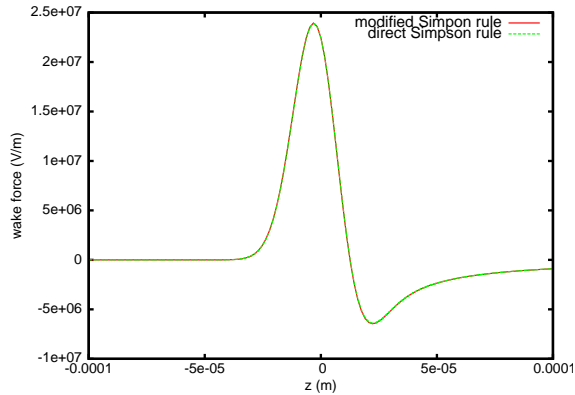}
\caption{
The CSR wakefield as a function of 
bunch length coordinate in an electron beam with the modified Simpson rule
($137$ grid points) and the direct Simpson rule ($8191$ grid points).}
\label{figcsrfldcmp}
\end{figure}

The second integral in the Eq.~\ref{csr} will be
calculated using an integrated Green's function method 
with constant current density basis
functions 
%This might be more effective than the direct numerical
%Simpson quadrature rule 
given the fact
that the CSR wake function varies sharply while
the line density changes slowly inside such a short-range interval.
At each grid point $z_i$, the interval $[z_i-\delta z,z_i]$ is 
subdivided into 
$N_{int}$ number of sub-intervals with a length 
$h_{int} = \delta z/(N_{int}-1)$.
This integral can now be rewritten as:
\begin{eqnarray}
\int_{z_i-\delta z}^{z_i} W(z_i-z') \rho(z') dz' & = &
\int_{z_1}^{z_1+h_{int}/2} W(z_i-z') \rho(z') dz' + \sum_{j=2}^{i-1} \int_{z_j-h_{int}/2}^{z_j+h_{int}/2} W(z_i-z') \rho(z') dz' \nonumber \\
& &
+\int_{z_i-h_{int}/2}^{z_i} W(z_i-z') \rho(z') dz'
\end{eqnarray}
where $z_1 = z_i - \delta z$.
Since the $\delta z$ is small (O($\mu m$)), only a small number of sub-intervals are needed. 
Assuming that the current density is constant within each sub-interval,
the above integral can be approximated as:
\begin{eqnarray}
\int_{z_i-\delta z}^{z_i} W(z_i-z') \rho(z') dz' & = &
\rho(z_1) \int_{z_1}^{z_1+h_{int}/2} W(z_i-z') dz'  + \sum_{j=2}^{i-1} \rho(z_j) \int_{z_j-h_{int}/2}^{z_j+h_{int}/2} W(z_i-z') dz' \nonumber
 \\
& & + \rho(z_i) \int_{z_i-h_{int}/2}^{z_i} W(z_i-z') dz'
\end{eqnarray}
The integral with respect to the wake function in above equation can be
done numerically if an analytical solution is not available. Fortunately,
in the case of the steady-state CSR wake function, 
there exists an analytical solution for the integral of the wake function~\cite{murphy},
i.e.
\begin{eqnarray}
\int_{z_j-h_{int}/2}^{z_j+h_{int}/2} W(z-z') dz' & = & [\nu(A(z-z_j+h_{int}/2))-
\nu(A(z-z_j-h_{int/2}))]/A
\end{eqnarray}
where
\begin{eqnarray}
\nu(u) & = & -\frac{e \gamma^4}{4 \pi \epsilon R^2}\frac{3}{4}\{-\frac{2}{u} +\frac{1}{u\sqrt{u^2+1}}(\Omega^{1/3}+
\Omega^{-1/3}) + \frac{2}{\sqrt{u^2+1}}(\Omega^{2/3}-\Omega^{-2/3})\}
\end{eqnarray}
and $\Omega = u + \sqrt{u^2+1}$, and $A = 3\gamma^3/(2R)$.
Using this modified Simpson rule, we recalculated the CSR wakefield 
inside the electron beam using $10$ sub-intervals inside the short range
around $0$ and $127$ grid points for the rest of the region. The results
are shown in Fig.~\ref{figcsrfldcmp} together with the 
direct Simpson rule calculation using $8191$ grid points.
It is seen that the wakefield from the modified Simpson rule agrees with
the high resolution direct Simpson rule very well even with only $137$ 
grid points.  

\section{Summary}

In this paper, we present a fast high-order method for
numerical calculation of the wakefield on electrons inside
an electron beam. Using the Simpson quadrature rule, we
present examples to calculate the wakefields from the
resonator wake function and from the
CSR wake function. These examples can have $O(h^4)$ accuracy 
with a $O(Nlog(N))$ computational cost. 
Besides the
fast speed and potential high numerical accuracy, the 
above CSR wakefield calculation also uses the direct line density instead of 
the first derivative of
the line density with numerical filtering of the density as suggested in 
Reference~\cite{borland}.
This paper gives application examples to the calculation of the longitudinal
wakefield for an electron beam. The same method can also be applied to
the calculation of the transverse wakefields inside an electron beam using
a modified density function. 

\section*{Acknowledgements}
This research was supported by the Office of Science of the U.S. Department of Energy under Contract No. DE-AC02-05CH11231.
This research used resources of the National Energy Research Scientific Computing Center.


\begin{thebibliography}{9}   % Use for  1-9  references
%\begin{thebibliography}{99} % Use for 10-99 references
\bibitem{qiang0} 
J. Qiang, R. D. Ryne, S. Habib, V. Decyk, J. Comp. Phys. vol. 163, 434, (2000).
\bibitem{qiang00} 
J. Qiang, R. D. Ryne, M. Venturini, A. A. Zholents, I. V. Pogorelov,
Phys. Rev. ST Accel. Beams, 12, 100702  (2009).
\bibitem{elegant} 
M. Borland, ANL Advanced Photon Source Report
No. LS-287, 2000.
\bibitem{lietrack} 
K. Bane, P. Emma, in Proc. of PAC05, p. 4266, 2005.
\bibitem{hockney}R. W. Hockney and J. W. Eastwood,  {\it Computer Simulation
Using Particles}, Adam Hilger: New York, 1988.
\bibitem{nr} W. H. Press, B. P. Flannery, S. A. Teukolsky, and W. T. Vetterling,{\it Numerical Recipes in FORTRAN: The Art of Scientific Computing}, 2nd ed. Cambridge, England: Cambridge University Press, 1992. 
\bibitem{qiang3}
J. Qiang, S. Lidia, R. D. Ryne, and C. Limborg-Deprey, Phys. Rev.
ST  Accel. Beams {\bf 9}, 044204 (2006).
\bibitem{ryne1} Robert D. Ryne, On FFT-based convolutions and correlations, with application to solving
Poisson's equation in an open rectangular pipe, arXiv:1111.4971 (2011).
\bibitem{qiang1}
J. Qiang, Comp. Phys. Comm. 18, 313316, (2010).
\bibitem{math}M. Abramowitz and I. A. Stegun, eds. {\it Handbook of Mathematical Functions with Formulas, Graphs, and Mathematical Tables}, New York: Dover, 1972. 
\bibitem{math2}http://mathworld.wolfram.com/Newton-CotesFormulas.html.
\bibitem{bane} K. L. F. Bane, ``Resistive Wall Wakefield in the LCLS Undulator Beam Pipe,'' SLAC-PUB-10707, Revised October 2004.
\bibitem{saldin}
E. L. Saldin, E. A. Schneidmiller, M. V. Yurkov, Nucl. Inst. Meth. Phys. Res. A 398, pp. 373-394, 1997.
\bibitem{ryne}
R. D. Ryne, B. Carlsten, J. Qiang, N. Yampolsky, 
``A Model for One-Dimensional Coherent Synchrotron Radiation
including Short-Range Effects,'' arXiv:1202.2409 (2012).
\bibitem{murphy}
J. B. Murphy, S. Krinsky, and R. L. Gluckstern, Particle Accelerators, Vol 57, pp. 9-64, 1997.
\bibitem{borland}M. Borland, Phys. Rev. Sepecial Topics - Accel. Beams 4, 070701 (2001).
\end{thebibliography}
\end{document}